# Phase stability of Ni–Al nanoparticles

S. Ramos de Debiaggi[1], J.M. Campillo[2], A. Caro[3]

[1]Departamento de Física, Universidad Nacional del Comahue, 8300 Neuquen, Argentina
[2]Elektrika eta Elektronika Saila, Euskal Herriko Unibertsitatea, 48080 Bilbo, Spain
[3]Centro Atómico Bariloche, 8400 Bariloche, Argentina

**Abstract**

The phase stability of Ni–Al clusters of nanometer size was studied by using the embedded atom model and Monte Carlo simulation techniques. For temperatures of 500 and 1000 K and for a range of compositions below 70 at.% Al, the equilibrium structures of the system were determined and compared with the bulk results. We found that the bulk NiAl (B2) and $Ni_3Al$ ($L1_2$) phases were stable phases in the nanoparticle system; however, for deviations from ideal composition, the analysis revealed that, because of the surface effect, the composition of the clusters was not uniform. There was a core region in which the structure was ordered, B2 or $L1_2$, with a composition very close to the ideal, and a chemically disordered mantle region that allocated the deviations from ideal stoichiometries; in this way, a larger phase field appeared, indicating trends similar to those found in experiments on nanocrystalline Ni–Al powder [S.K. Pabi and B.S. Murty, Mater. Sci. Eng. A214, 146 (1996)]. For concentrations between 37 and 51 at.% Al, an intermediate phase, similar to the tetragonal $L1_0$ martensite, appeared.

## 1. Introduction

Because of their excellent high-temperature mechanical properties, a large amount of research has been performed on Ni–Al intermetallic compounds. At high temperatures (T ≈ 1000 K), the Ni–Al phase diagram indicates the existence of a Ni-rich solid solution phase, an ordered $L1_2$ phase for Ni concentrations around 75 at.%, and an ordered B2 phase for Ni concentrations around 50 at.% Al. At lower temperatures, there are various other phases that can be found in the Ni–Al system. The Ni-rich Ni–Al alloys with Ni contents greater than about 61 at.% transform martensitically upon cooling from elevated temperatures to give the $L1_0$ structure based on the face-centered-tetragonal (fct) lattice. Two possible structures are formed depending on the initial Ni content: ABC (3R) stacking or ABCABAC (7R) stacking. Ni–Al alloys with a Ni content greater than 63 at.% transform to 3R martensite, whereas alloys with less than 63 at.% Ni content transform to 7R martensite. The tetragonal shape of these structures can be understood in terms of the volume ratio between the Al and Ni atoms; in the $L1_0$ structure, Al atoms which are larger than Ni, are placed as nearest neighbors in an *xy* plane, thus increasing the lattice parameter in that plane. The Ni-rich compositions, when aged at temperatures greater than about 973 K, transform to two-phase $L1_2$ + B2 microstructures. A stable phase with a stoichiometry of $Ni_5Al_3$ exists between 62.5 and 68 at.% Ni around 773 K. This $Ni_5Al_3$ has an orthorhombic crystal structure; its phase boundary locations and lattice parameters have been investigated.[1] In addition, a short-range ordering reaction can occur at lower temperatures to form a metastable $Ni_2Al$ phase.

The phases occurring in the Ni-rich NiAl alloys are related among themselves when viewed in their tetragonal configuration.[2] The B2 structure has an associate tetragonal $L1_0$ structure, that can be seen rotating the lattice vectors in the *xy* plane along the *xy* face. The new lattice parameters are $a' = b' = \sqrt{2}a$ and $c' = a$, where $a'$ is the lattice parameter of the cubic B2 phase. The axial ratio is then

$c/a = 0.71$. So, starting with the B2 NiAl alloy and increasing the Ni content, a progressive substitution and ordering of the Ni atoms on the Al sublattice takes place with associated changes in the axial ratio in the case of stable phases NiAl, $Ni_5Al_3$, and $Ni_3Al$. In the martensite $L1_0$ the atomic arrangement of the atoms is essentially the same as that for the B2 NiAl, with excess Ni atoms substituting on the Al sublattice and with a different axial ratio due to a Bain distortion.

In recent years, nanostructured materials have been the focus of great interest.[3,4] The reasons are the different behaviour they show compared with bulk coarse polycrystalline materials and their potentially useful properties.[5–9]

Ni–Al nanocrystalline materials can be synthesized by high-energy ball milling or mechanical alloying (MA) of elemental or intermetallic compound powders. Using MA it has been possible to obtain nanocrystallites in equilibrium phases.[10,11] Other nonequilibrium states, such as amorphous phases, can also be obtained.[12,13] There has been considerable interest in the study of the structural properties of the samples obtained in this way as a function of milling time and average grain size. The formation of disordered nanocrystalline facecentered-cubic (fcc) $Ni_3Al$ by MA of elemental powders has been followed by x-ray diffraction (XRD) and differential scanning calorimetry (DSC) in samples with average grain sizes around 10 nm. An exothermic ordering transition from fcc $Ni_3Al$ to $L1_2$ $Ni_3Al$ was found at 638 K.[10]

More recently, Pabi and Murty[11] have investigated $Al_3Ni$, NiAl, and $Ni_3Al$ nanocrystalline powders. They found that $Al_3Ni$ and NiAl were always ordered, and $Ni_3Al$ was disordered. The low ordering energy (5 kJ/ mol) appears to be the reason to obtain $Ni_3Al$ in a disordered state. Their results also indicate a large extension of the NiAl and $Ni_3Al$ phase fields, particularly toward the Al-rich compositions.

In bulk nanostructured materials a large fraction of atoms are located at interfaces. This fact plays a role in determining the overall properties. Similarly, the surface plays an important role in isolated nanoparticles. On the one hand, the surface energy determines the equilibrium shape of the particle; on the other hand, and in general terms, it is energetically favorable to create a defect on the surface, instead of an antisite defect or a vacancy in the bulk. It is thus expected that stoichiometry departures will be accommodated differently in nanoparticles than in bulk specimens, with consequences on the extensions of phase fields.

In this work we used computer simulation techniques to study the structural properties of Ni–Al nanoparticles of very small size (2.8 nm in diameter), chosen to highlight the surface effects. For a range of Al concentrations below 70 at.%, the predicted equilibrium structures at temperatures of 500 and 1000 K are analyzed in terms of the cluster composition. They are then compared with the known phase diagram of bulk Ni–Al alloys and with recent experimental information about a closely related problem, nanocrystalline Ni–Al powder.[11]

## 2. Computational procedure

The many-body embedded atom model (EAM), developed by Daw and Baskes,[14] is an efficient and relatively reliable method for representing the atomic interactions in fcc metals and their alloys. In this work, we have used the empirical EAM potentials developed by Foiles and Daw[15] to describe the Ni–Al system. These EAM functions were determined by fitting to the equation of state (equilibrium lattice constant, sublimation energy, and bulk modulus are then exactly reproduced), elastic constants and vacancy formation energy of the pure metals Ni and Al and to the sublimation energy of the ordered alloys NiAl (B2 phase) and $Ni_3Al$ ($L1_2$ phase). In agreement with the experimental data for the bulk Ni–Al system at 1000 K, this model predicts[15] the existence of a Ni-rich fcc solid solution phase, an ordered L1$_2$ phase for Ni concentrations around 75 %, and an ordered B2 phase for Ni concentrations around 50 %. Point defect properties for the Ni3Al phase are also reported in that work.

Other EAM potentials for the Ni–Al system are available in the literature. Voter and Chen[16] developed a potential to describe the Ni$_3$Al phase, and Mishin and Farkas[17] extended it to account for NiAl and Ni$_5$Al$_3$ phases. There is no consensus on a single potential able to describe all point defect properties in these phases.[18] Our choice of the Foiles and Daw[15] potential relies on the reported stability of the phases that interest us.

To analyze the nanoparticle phase stability, we generated a Ni–Al spherical cluster of 2.8 nm containing 959 atoms, whose initial configuration is not relevant because the equilibrium atomic structure and composition of this Ni–Al nanocrystal is simulated by Monte Carlo (MC) techniques. The calculations were performed in an isobaric, isothermal, transmutation ensemble, where the total number of particles and the relative chemical potentials of the two species were held fixed (NPT ensemble with transmutations). The pressure was set to zero, and the temperatures used in the simulations were 500 and 1000 K.

For a given temperature and pressure, the equilibrium composition and structure were determined as a function of the relative chemical potential of the two elements $\Delta \mu = (\mu_{Al} - \mu_{Ni})$. In the bulk Ni–Al samples, these curves provide information about the existence of stable single phases as well as the range of homogeneity of each phase, defined by the compositions at which the structure changes as a function of chemical potentials.

For each value of the relative chemical potential, the system evolves with MC steps, which can be either a displacement or a change in chemical identity. After an equilibration stage, average quantities were obtained during $10^6$ MC steps.

To analyze the atomic structure of the resulting nanocrystalline Ni–Al samples, the radial distribution function and three-dimensional plots are presented for some representative configurations extracted from the sequence of a million steps.

## 3. Results and discussion

Figure 1 shows the results for the equilibrium concentrations as a function of the relative chemical potential at temperatures of 500 and 1000 K and zero pressure together with those obtained for the bulk material.[15] For the bulk material at 1000 K, the computed phase diagram for the Ni-rich region indicates the existence of three equilibrium structures, in agreement with the experimental phase diagram. The first phase corresponds to NiAl, with B2 structure, the second is Ni$_3$Al, with L1$_2$ structure, and the third is the Ni-rich solid solution with fcc structure. These results are obtained from the MC simulation as follows (see Fig. 1): the bulk system in a computer simulation is characterized by single-phase regions, where the composition varies continuously with $\Delta \mu$. The boundaries of these single-phase regions show discontinuities in the composition as a function of $\Delta \mu$. For large and negative $\Delta \mu$, a fcc solution is obtained. A plateau in the composition starting at $\Delta \mu \approx -0.5$ eV indicates the transition from a fcc solid solution to the ordered L1$_2$ phase. If $\Delta \mu$ is further increased, the equilibrium structure is still the ordered L1$_2$ phase with structural defects that accommodate the increasing amounts of Al. For $\Delta \mu \approx 0.5$ eV there is a large jump in the composition, indicating transition to the B2 structure.

This model predicts the range of homogeneity of these phases as follows: the fcc solid solution extends from pure Ni up to 18 at.% Al. The L1$_2$ phase extends from 22 to 34 at.% Al. The B2 phase is found for concentrations higher than 42 at.% Al. These phase boundaries are in reasonable agreement with experiments, although for the fcc solid solution and the L1$_2$ phases are somewhat wider than those found experimentally. This fact certainly has a contribution from finite size effects in the computer simulations that preclude the existence of inhomogeneous configurations corresponding to two-phase fields. Therefore, the phase field extensions determined for the bulk system do not necessarily reflect limitations of the potential but of the simulation technique for infinite systems.

In nanoparticles a different behavior is observed. We do not see regions of constant composition, characteristic of the phase stability in the bulk, or discontinuities in the composition, which indicate the existence of phase transitions. At 500 K the shape of the curve still has reminiscences of the bulk, but at 1000 K this is completely lost, as it becomes an almost straight line (see Fig. 1). The disappearance of the typical plateau indicates that the composition varies continuously with the chemical potential. However, it does not imply that stable equilibrium phases (B2, L1$_2$, or fcc) do not exist in the nanocrystalline sample. In fact, analysis of the structures at 500 K for several compositions indicates that the structures remain close to the bulk equilibrium phases, even in those cases with large deviations from the ideal stoichiometries. As we discuss below, the difference with respect to the bulk behavior resides in the ability of the nanoparticle to accommodate on the cluster surface the structural defects introduced by the stoichiometry deviations.

This point is analyzed in detail by using the pair distribution functions of Al and Ni atoms, which allows us to deduce the crystal structure of the cluster at each composition. The Al (Ni) pair distribution functions have been obtained by using as central atom the Al (Ni) atom nearest the geometric center. The pair distribution functions we show are the averages over 10 different configurations representative of the equilibrium ensemble, taken with an interval of $10^5$ steps between each other. The resulting distribution functions at $T = 500$ K are shown in Figs. 2 and 3 for different compositions. The pair distribution functions for Al and Ni atoms in the perfect L1$_2$ and B2 structures are also shown for comparison.

If we consider the evolution of the structural properties of the clusters for decreasing values of $\Delta\mu$ and starting with $\Delta\mu = 1.5$ eV, we have the following results (see Figs. 2 and 3):

(i) For Al concentrations between 65 and 52 at.%, the distribution of peaks in the plots is very similar to the B2 structure, with a core region ($R \leq 8$ Å) in which almost all the peaks of the perfect B2 cluster are reproduced. A mantle in which a superposition of peaks is observed surrounds this core. The excess Al with respect to the ideal stoichiometry (50 at.%) is located on this mantle, and the core has almost perfect stoichiometry. Therefore, the cluster composition is not uniform. Note this segregation of Al atoms to the surface clearly reflected in the extra Al peak for $R \approx 14$–16 Å (see Fig. 3), whereas in the perfect B2 cluster there are no contributions of Al atoms at those distances. A three-dimensional plot of the cluster shows that the shape is an elongated sphere. As the Al concentration decreases, approaching the ideal value for the B2 structure, the extra Al outer peak moves inward, the ellipsoid deforms toward a sphere, and the cluster reduces its volume.

(ii) There is an interesting transition region between 51 and 37 at.% Al, in which the structures cannot be related to equilibrium bulk phases. In fact, a transition from B2 to L12 seems to occur through an intermediate phase. At 45 at.% Al, for example, it is observed that the distribution of the Ni atoms adopts an inner core ($R < 6$ Å) in which the positions of the peaks correspond to compact packaging, although their relative heights do not correspond to L1$_2$ or fcc. Visual inspection of the core structure reveals that it is closely related to the tetragonal face-centered structure L1$_0$ (or TP4), which is the martensitic phase of the Ni–Al system. The lattice parameters of the tetragonal core phase can be estimated (with some error due to thermal fluctuations); they are $a = b = 3.8$ Å and $c = 3.4$ Å, with an axial ratio of about 0.9. A similar situation appears for a concentration of 40 at.% Al; for lower chemical potential difference, the structures evolve toward the L1$_2$ phase.

(iii) Between 35 and 21 at.% Al, the nanocluster has a structure very similar to the L1$_2$ phase. In Figs. 2 and 3 the pair distribution functions for 33 at.% Al are shown. There is a core region ($R \approx 12$ Å) with almost the ideal 25 at.% Al concentration; the mantle is chemically disordered and allocates the excess of Al present.

(iv) For a concentration of 17 at.% Al and below, the peaks in the Al and Ni distribution functions are located according to an fcc structure, but their relative intensities do not match the L1$_2$ distribution function; the structure is a substitutional solid solution of Al in fcc Ni. However, the absence of the nearest neighbor peak in the Al distribution function reveals some short-range order in the chemically disordered solid solution.

In summary, for $T = 500$ K we can clearly identify in the plot of the composition versus relative chemical potential three of the bulk phases: the solid solution and the L1$_2$ and B2 phases. However, the structures are not perfect crystallites but show a highly ordered core surrounded by a chemically disordered mantle, where most of the deviations from ideal stoichiometries are allocated. The dimensions of the core do not change appreciably with composition; they are approximately $R \approx 8$ Å for the B2 phase and $R \approx 12$ Å for the L1$_2$ phase.

In rigorous terms, the range of existence of the pure phases, which in the bulk means the crystal plus structural defects that accommodate the departures from ideal stoichiometry, is reduced in nanoparticles if we look at the core composition, because the surface accommodates the defects, producing almost defect-free cores. What is then relevant to analyze is the composition range in which a core with one almost defect-free phase, and a cortex with a surface phase, coexist. In this sense we show in Fig. 4 the range of existence of a given core phase for particles of 2.8 nm at 500 K. It is important to point out that any possible phase field extension then has this particular interpretation because of the nonuniformity of the cluster composition. The phase boundaries are approximate, representing the concentration values obtained in the particular set of simulations done. We conclude from this figure that, at 500 K:

(i) The fcc solid solution is wider than the experimental bulk value (up to about 17 instead of 7 at.% Al).

(ii) The L1$_2$ phase exists in the range 21–35 at.% Al, instead of 24–26 at.% in the bulk.

(iii) The region between 37 and 51 at.% Al is a tetragonal phase close to the martensitic $L1_0$ (3R), apparently a characteristic phase of the nanoparticles, as predicted by this model.

(iv) The B2 phase is found for concentrations higher than 52 at.% Al, a lower limit that is shifted with respect to the bulk (42 at.%). The upper limit has not been determined.

At $T = 1000$ K the structural behavior of the Ni–Al nanoparticles follow approximately the same trends as for $T = 500$ K. However, because of the higher temperature the structures found are substantially more disordered, and stronger Al surface segregation effects are observed. Even though phase fields are less clearly defined at this temperature, an fcc solid solution phase is found up to 26 at.% Al; between 28 and 34 at.% Al we found a very disordered $L1_2$ core phase; and the B2 core phase is found for concentrations higher than 52 at.% Al. The lower stability range of the $L1_2$ phase with respect to $T = 500$ K and the fact that the structure is very disordered is to be expected considering the low ordering energy.

A transition region, with more disordered structures is found between 35 and 52 at.% Al. As for $T = 500$ K, for Al concentrations around 42–48 at.% Al, a tetragonal $L1_0$ core phase coexisting with an outer disordered shell can be distinguished.

## 4. Summary and conclusions

Pabi and Murty[11] have recently published results on phase fields on ball-milled nanoparticles with grain sizes larger than, but close to, those considered here ($d > 6$ nm). The characteristics of the system they studied are closely related to ours; however, two significant differences must be pointed out. First, the surface-to-volume ratio of our particles is twice as large as theirs; therefore, the surface effects we observe are enhanced in our study. Second, we model free particles, and the experimental setup deals with a powder in which a mixture of free surfaces as well as interfaces appear. Therefore, comparison of our results and the experimental report can be done only qualitatively.

The disordered $Ni_3Al$ and order NiAl and $Al_3Ni$ phases are found by Pabi and Murty[11] in a range of compositions between 10–30, 35–75, and 75–78 at.% Al, respectively. An extension of the $Ni_3Al$ range of stability is found and, in the NiAl nanocrystalline phase, the range of stability extends more toward the Al-rich side than to the Ni-rich side. The last result has been interpreted in terms of the additional tolerance the conventional NiAl structure presents to vacancies in the Ni sublattice with respect to the vacancies in the Al sublattice. However, another ingredient that emerges from our calculations should be taken into account. Because of the presence of the surface, the compositions of the clusters may be not uniform, even at these larger sizes, indicating that the phase field modifications found for both NiAl and $Ni_3Al$ phases may have a contribution coming from the effects reported here.

Whether the tetragonal $L1_0$ phase is a new phase characteristic of the Ni–Al nanoparticle cannot be answered definitely because this is a model system based on a semiempirical approximation. Other potentials, like those reported recently,[18] which were specifically designed to describe the properties of NiAl and the orthorhombic $Ni_5Al_3$ phases, may be used to support this possibility. It is interesting to note that one can image this tetragonal phase with $c/a = 0.9$ as an intermediate structure between the B2 ($c/a = 0.71$) and the $L1_2$ ($c/a = 1$) structures.

In summary, the structural properties of 2.8-nm Ni–Al nanocrystalline particles have been analyzed by MC techniques. For the range of compositions considered, and for $T = 500$ K, we can clearly distinguish the three phases also present in the bulk: B2, L12 and the fcc solid solution, and a tetragonal phase similar to the $L1_0$ martensite. For both NiAl and $Ni_3Al$ nanocrystalline phases, the crystallites are found to have an ordered core with compositions close to the ideal stoichiometry and a chemically disordered mantle in which larger deviations from the ideal compositions can be accommodated. The apparent modifications in the phase field extensions can be related to the nonuniformity of the cluster composition generated by the Al surface segregation effect.

**Ackowledgments**

The authors are grateful to G. Aurelio and J. Silva Valencia for their contributions to the initial phase of this work and to A. Fernandez Guillermet for his comments and suggestions. One of us (J.M.C.) acknowledges financial support from the Universidad Nacional del Comahue and the Intercampus Program. This work was partly supported by a U.N. Comahue Grant.

# Figures

**Figure 1.** Calculated equilibrium Al concentration as a function of the relative chemical potential ($\mu_{Al} - \mu_{Ni}$) for zero pressure. Dashed line is the bulk equilibrium alloy concentration at $T = 1000$ K taken from Ref. 15. Triangles correspond to the Ni–Al nanosphere at 500 K. Squares are results for $T = 1000$ K. For clarity, triangles and squares have been shifted in the vertical axes by 10 % and 20 %, respectively.

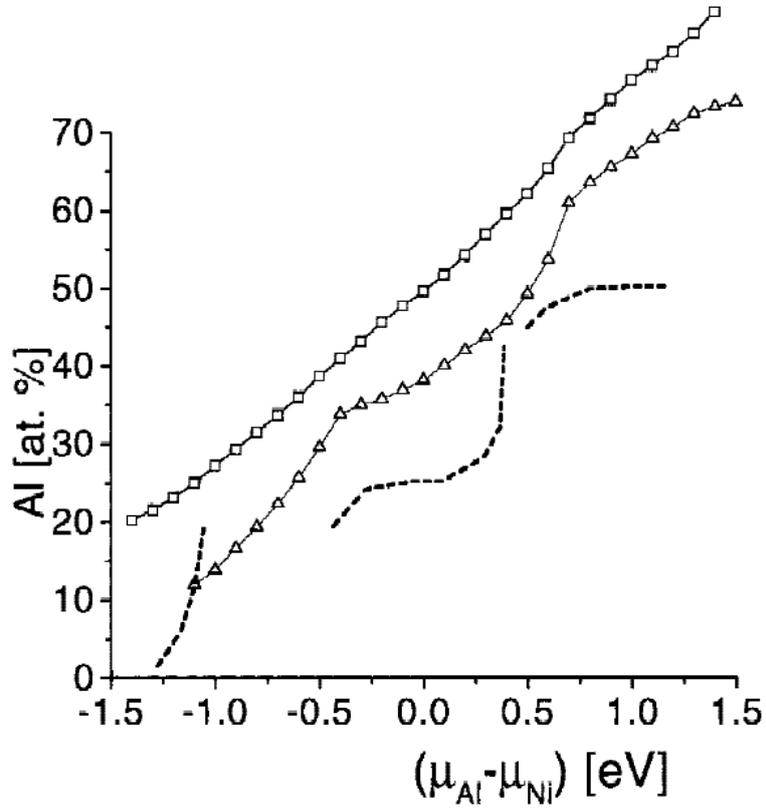

**Figure 2.** Ni pair distribution functions at different equilibrium compositions for the Ni–Al nanosphere at $T$ = 500 K. Perfect $L1_2$ and B2 functions are included for comparison.

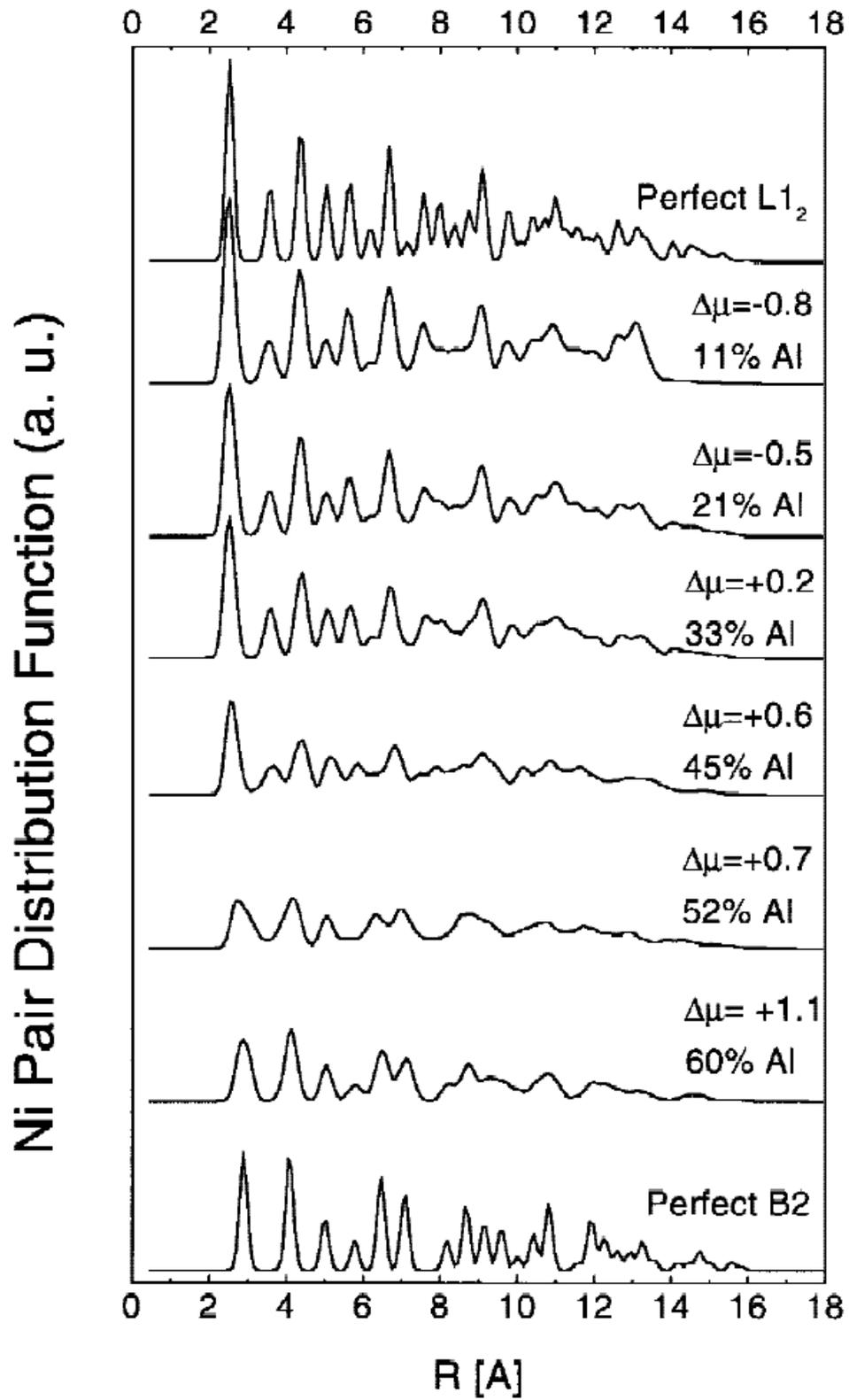

**Figure 3.** Al pair distribution functions at different equilibrium compositions for the Ni–Al nanosphere at $T = 500$ K. Perfect L1$_2$ and B2 functions are included for comparison.

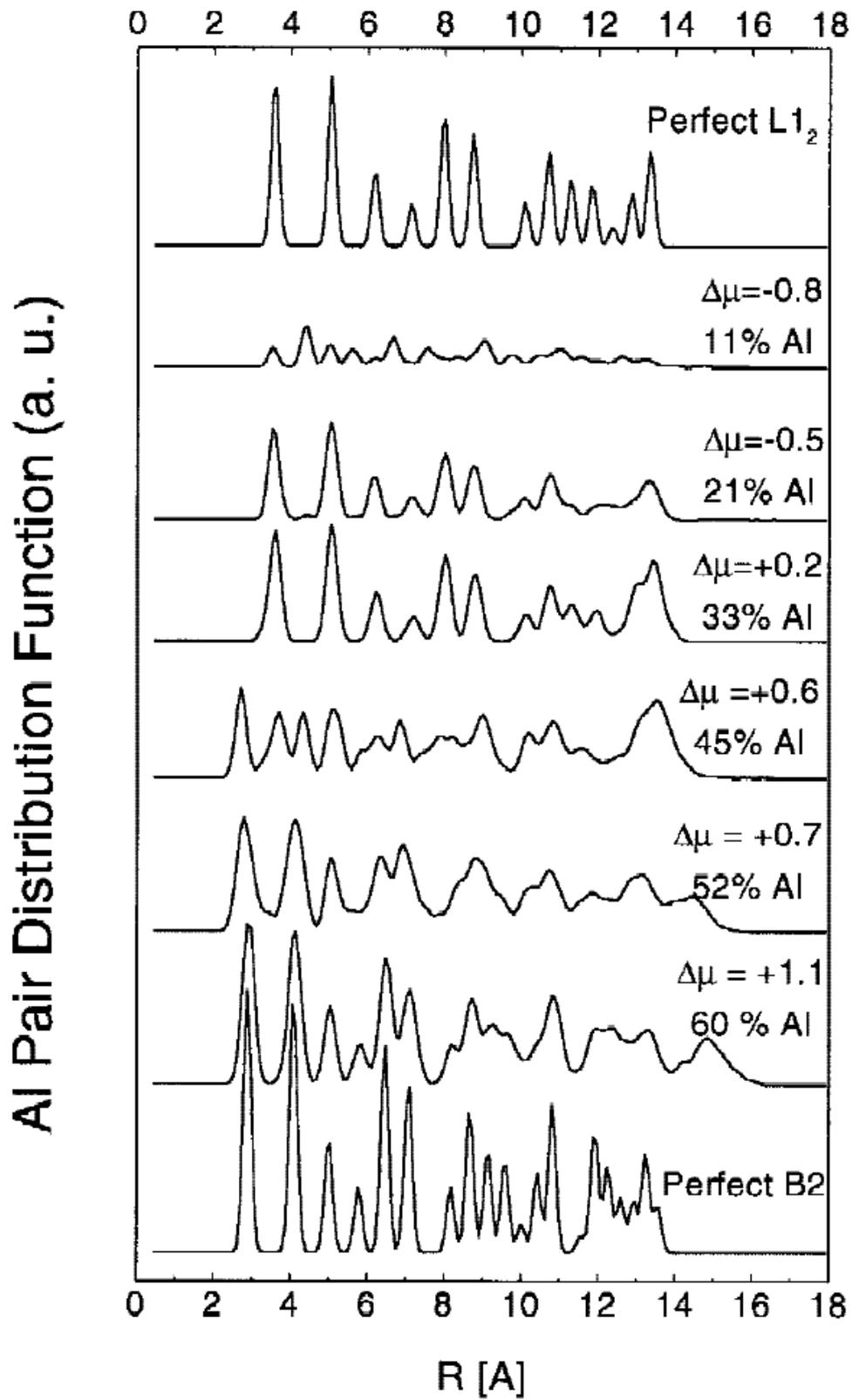

**Figure 4.** Schematic range of existence of different phases for bulk experimental at 500 K (upper), nanoparticles of 2.8 nm as predicted in this work (middle), nanoparticle experimental after 30 h of mechanical alloying[11] (lower). Phase limits in the middle graph are approximate, representing the concentration values obtained in the particular discrete set of simulations done. In particular, the low Ni limit of the B2 phase has not been determined.

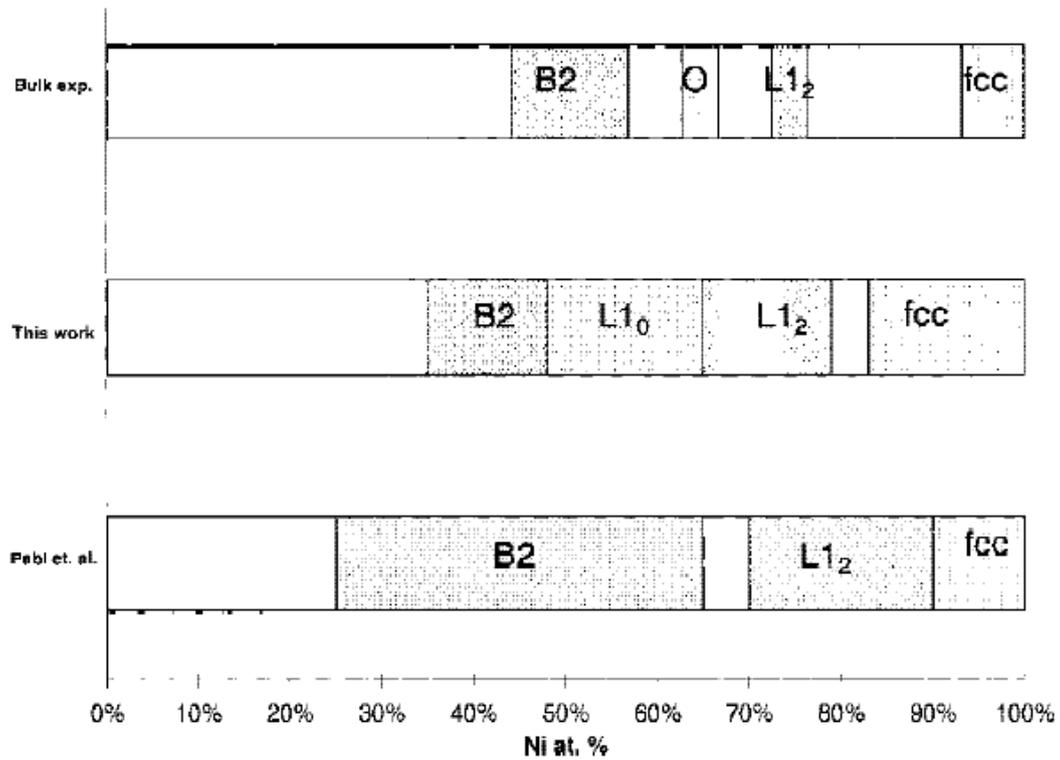